\documentclass[lettersize,journal]{IEEEtran}
\usepackage{amsmath,amsfonts}
\usepackage{algorithmic}
\usepackage{algorithm}
\usepackage{array}
\usepackage[caption=false,font=normalsize,labelfont=sf,textfont=sf]{subfig}
\usepackage{textcomp}
\usepackage{stfloats}
\usepackage{url}
\usepackage{verbatim}
\usepackage{graphicx}
\usepackage{cite}

\usepackage{amssymb,bbm} 
\usepackage{multirow} 
\usepackage{booktabs} 
\usepackage{xcolor}
\usepackage{siunitx}

\definecolor{darkgreen}{rgb}{0.0, 0.5, 0.0}
\definecolor{darkorange}{rgb}{0.5, 0.3, 0.1}
\definecolor{orange}{rgb}{1,0.5,0}

\newcommand{\eco}[1]{\textcolor{black}{#1}}

\newcommand{\ntx}
{N_{\text{Tx}}}

\title{Belief-Adaptive MAP Detection for Molecular ISI Channels with Heteroscedastic Noise}

\author{Erencem Ozbey, H. Birkan Yilmaz, and Chan-Byoung Chae

    \thanks{E. Ozbey and H. B. Yilmaz are with NETLAB, Department of Computer Engineering, Bogazici University, Istanbul, Turkiye. (e-mails:ozbeyerencem@gmail.com; birkan.yilmaz@bogazici.edu.tr)} 
    \thanks{C.-B. Chae is with the School of Integrated Technology, Yonsei University, Seoul 03722 Korea (e-mail: cbchae@yonsei.ac.kr)}
}

\begin{document}

\maketitle

\begin{abstract}
Inter-symbol interference (ISI) with heteroscedastic (state-dependent) noise is a defining feature of molecular communication via diffusion (MCvD). However, such noise variance dependency across ISI states has not been systematically considered in prior detector designs. This letter introduces two decoding mechanisms, Belief-Adaptive Maximum A Posteriori (BA-MAP) and Soft BA-MAP, that explicitly incorporate state-dependent count means and variances of the molecular channel. The BA-MAP method derives per-symbol adaptive MAP thresholds based on the receiver's current state beliefs, whereas Soft BA-MAP computes mixture log-likelihood ratios by weighting all possible ISI states. Simulation and analyses confirm that the proposed detectors outperform conventional equalization and fixed-threshold methods, \eco{and approach ideal zero-decision-delay MAP detection with perfect ISI-state knowledge.}

\end{abstract}

%
%%%%%%%%%%%%%%%%%%%%%%%%%%%%%%%%%%%%%%%%%%%%%%%%%
%%%%%%%%%%%%%%%%%%%%%%%%%%%%%%%%%%%%%%%%%%%%%%%%%
%%%%%%%%%%%%%%%%%%%%%%%%%%%%%%%%%%%%%%%%%%%%%%%%%
%%%%%%%%%%%%%%%%%%%%%%%%%%%%%%%%%%%%%%%%%%%%%%%%%
%
\section{Introduction}
\IEEEPARstart{M}{olecular} communication is a paradigm in which molecules are used as information carriers instead of electromagnetic waves or electrical signals~\cite{birkan2016survey,chae2023iobnt}.
Inspired by biological systems, molecular communication via diffusion (MCvD) encodes information into molecules that are released into a diffusive medium by a transmitter (Tx) node to convey information. The receiver (Rx) detects these molecules and decodes the transmitted information ~\cite{channel_modeling}.

In general, MCvD systems operate with fixed-duration symbol intervals. However, due to the long diffusive tail, molecules released during one interval may continue to arrive in several subsequent intervals, leading to severe inter-symbol interference (ISI). Several modulation techniques have been proposed to mitigate ISI. For example, Molecular Shift Keying (MoSK)\cite{modulations_kuran}, where the Tx emits different types of molecules, and Zebra-CSK\cite{zebra-csk}, which uses inhibitors to destroy molecules in the environment, are modulation-level approaches designed to reduce ISI.
A modulation scheme defines how information is encoded in one or more properties of molecular signals (e.g., molecule type, concentration, or release timing), which in turn influences the effective channel impulse response. However, the impulse response itself does not solely determine communication performance, reliable transmission also depends on the detection strategy employed at the Rx~\cite{detection_theory}.

In MCvD systems, the received signal is generally represented by the number of molecules observed as a function of time.
The arrival of molecules in an interval follows a binomial model ~\cite{yilmaz2014arrival}. Under common counting models (Poisson and Gaussian approximations), both the mean and the variance of the received molecules in an interval depend on the recent bit history ~\cite{damrath2017equivalent}. Thus, the channel is a finite–state channel with memory and heteroscedastic (state-dependent) noise. Simple fixed threshold detectors disregard this memory and variance structure. In order to mitigate ISI under a modulation, one alternative is to use Tx-side equalization techniques such as Power Adjustment (PA)~\cite{burcu-MTSK}.  On the other hand, Rx-side equalizers such as MMSE~\cite{receiver_akan} reduce effective memory but typically revert to a single hard threshold after linear processing, leaving variance asymmetry underexploited.

\eco{Belief-based and LLR-based detection principles are well established for finite-state communication channels, e.g., in sequence detection and forward-recursion-based MAP receivers~\cite{bcjr}. However, most MCvD detectors do not directly combine causal uncertainty over ISI states with state-dependent molecular count variance.} A classic way to account for channel memory is MAP sequence detection~\cite{receiver_akan} using the Viterbi algorithm~\cite{viterbi}. However, this algorithm introduces a decision traceback delay and requires additional storage for survivor paths. Blockwise iterative methods as introduced in~\cite{iterative-block} also introduce decision delay and are not suitable for time-critical scenarios. \eco{To address this gap, we propose two state-dependent signal detection methods that explicitly model heteroscedastic noise while operating with zero decision delay}:
\begin{itemize}
  \item \textbf{Belief-adaptive MAP thresholding (BA-MAP):} To obtain a lightweight detector, we approximate the state uncertainty using belief-matched Gaussians and select the threshold $\tau$ by the MAP condition.
  \item \textbf{Soft-belief, mixture MAP-LLR detection (Soft BA-MAP):} The Rx maintains a soft belief (a probability distribution) over ISI states. For each observation, it evaluates the posterior odds using belief-weighted Gaussian mixtures and decides by the sign of the log-likelihood ratio (LLR). This rule is Bayes-optimal for \eco{zero-delay} symbol-wise detection given the current state beliefs.
  \end{itemize}
\paragraph*{Contributions}
\begin{enumerate}
  \item We formulate the finite–state, heteroscedastic channel and derive a belief-driven \emph{mixture MAP--LLR} detector (Soft BA-MAP) that utilizes the full state mixture.
  \item We develop a \emph{belief-adaptive MAP threshold} (BA-MAP) that retains state awareness.
  \item We provide an information-rate framework and analyze the information rates of these strategies. %with existing equalization methods.
\end{enumerate}

\begin{figure*}[ht]
    \centering
    \includegraphics[width=1 \linewidth]{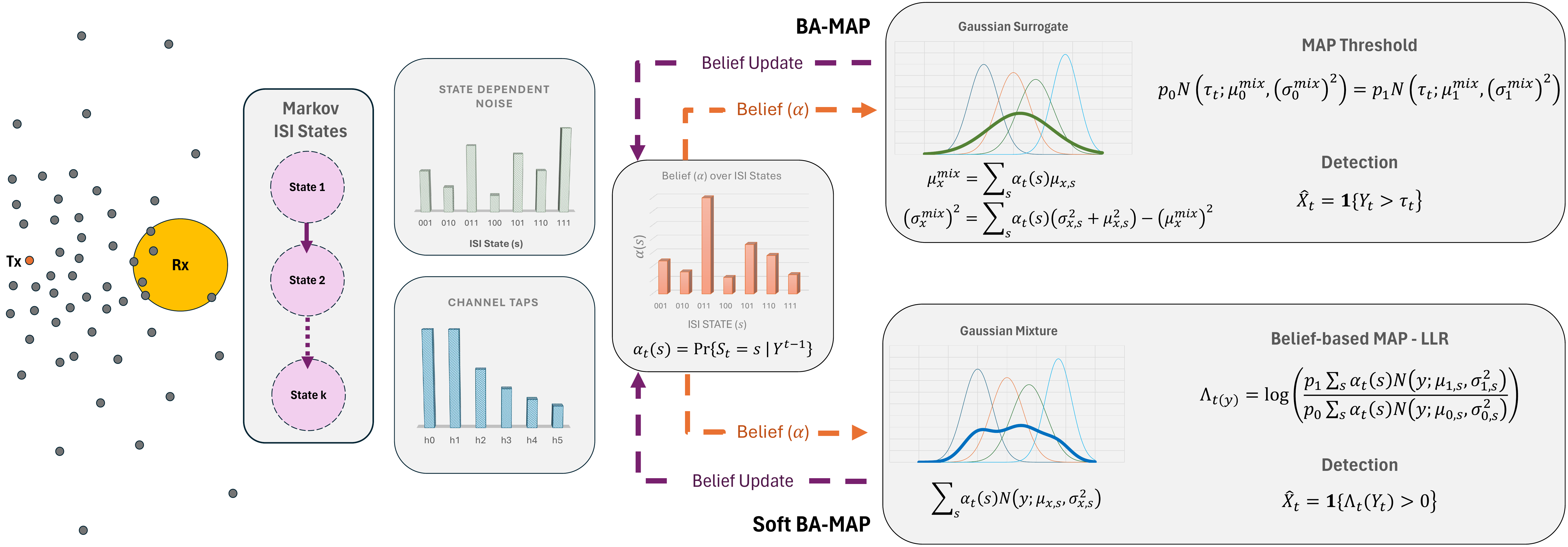}
    \caption{Belief-adaptive detection for molecular ISI channels with state-dependent noise. The MCvD channel is modeled as a finite-state ISI channel whose received-count statistics %(mean and variance) 
    depend on the ISI state. The Rx maintains a causal belief over ISI states via forward recursion, and uses this belief to adapt detection. Soft BA-MAP performs symbol-wise MAP detection using belief-weighted Gaussian-mixture likelihoods, whereas BA-MAP reduces the mixture to a single belief-adaptive Gaussian proxy via moment matching and applies a lightweight MAP threshold rule.}
    \label{fig:bam_overview}
\end{figure*}

%    
%%%%%%%%%%%%%%%%%%%%%%%%%%%%%%%%%%%%%%%%%%%%%%%%%
%%%%%%%%%%%%%%%%%%%%%%%%%%%%%%%%%%%%%%%%%%%%%%%%%
%%%%%%%%%%%%%%%%%%%%%%%%%%%%%%%%%%%%%%%%%%%
%%%%%%%%%%%%%%%%%%%%%%%%%%%%%%%%%%%%%%%%%%%%%%%%%
%%%%%%%%%%%%%%%%%%%%%%%%%%%%%%%%%%%%%%%%%%%%%%%%%
%%%%%%%%%%%%%%%%%%%%%%%%%%%%%%%%%%%%%%%%%%
%%%%%%%%%%%%%%%%%%%%%%%%%%%%%%%%%%%%%%%%%%%%%%%%%
\section{System Model}
\label{section_system_model}

We an MCvD system with \eco{an absorbing Rx} where communication proceeds in symbol intervals of duration $T_s$. For our channel formulation, $X_t\in\{0,1\}$ denotes the $t$-th symbol with prior $\Pr\{X_t=x\}=p_x$, i.e., $p_0$ represents the probability of sending bit-0. We consider a \eco{binary on-off keying} at the Tx side, i.e., when $X_t=1$, the Tx releases $\ntx$ molecules at the start of interval $t$ and if $X_t=0$, it releases none. At the Rx side, the number of absorbed molecules in the same symbol duration is denoted as $Y_t$.

Molecules from emissions in previous intervals arrive over future intervals and causes ISI. When the ISI memory order is $m$, we model the signal with $K\!=\!m{+}1$ taps:
\begin{align}
h_k \triangleq F_{\text{hit}}\!\big((k{+}1)T_s\big) - F_{\text{hit}}\!\big(kT_s\big),
\quad k = 0,1,\ldots,m
\end{align}
which are the arrival fractions in each symbol window. \eco{For the Rx, the number of molecules detected from each emission follows a Binomial distribution with success probability $h_k$, hence the corresponding per-molecule variance $v_k$ is $h_k(1-h_k)$.}
\eco{In our experiments, we choose memory order $m$ as the smallest memory order for which ($\sum_{k=0}^{m} h_k$) captures (75\%) of the total arrival probability.}
Conditioned on the recent $m{+}1$ bits $(X_t,\ldots,X_{t-m})$, the count $Y_t$ is modeled as
\begin{align}
\mu_t &= \ntx\sum_{k=0}^{m} h_k\,X_{t-k},\\
\sigma_t^2 &= \ntx\sum_{k=0}^{m} v_k\,X_{t-k},
\end{align}
%Per symbol, $Y_t$ is a sum of many independent Bernoulli trials contributed by current and past releases. When the expected count is moderate, the Central Limit Theorem justifies the Gaussian approximation
For each symbol, $Y_t$ is the \eco{sum of independent Bernoulli trials} from both the current and past emissions. When the expected molecule count is moderate, the Central Limit Theorem justifies approximating $Y_t$ as Gaussian
$Y_t \sim \mathcal{N}\!\big(\mu_t,\;\sigma_t^2\big).$
\eco{Accordingly, the detectors evaluated in this paper use Gaussian likelihoods for the molecular count observations.} We define an ISI state as the $m$-bit history $S_t=(X_{t-1},\ldots,X_{t-m})\in\{0,1\}^m$. The state space has $S=2^m$ elements and a deterministic next-state function \texttt{nxt}[s,x]. For each $(x,s)$ we precompute
\begin{align}
\begin{split}
\mu_{x,s} 
&= \ntx\!\Big(h_0 x+\sum_{k=1}^{m} h_k\,b_k(s)\Big),  \\
\sigma^2_{x,s} 
&= \ntx\!\Big(v_0 x+\sum_{k=1}^{m} v_k\,b_k(s)\Big),
\end{split}
\end{align}
where $b_k(s)\in\{0,1\}$ is the $k$-th bit of state $s$ \eco{corresponding to $X_{t-k}$}. Since we assume $Y_t$ is a Gaussian with state and input dependent mean/variance,
\[
Y_t \mid (X_t{=}x,S_t{=}s) \;\sim\; \mathcal{N}\!\big(\mu_{x,s},\,\sigma^2_{x,s}\big).
\]

%%%%%%%%%%%%%%%%%%%%%%%%%%%%%%%%%%%%%%%%%%%%%%%%%
%%%%%%%%%%%%%%%%%%%%%%%%%%%%%%%%%%%%%%%%%%%%%%%%%
%%%%%%%%%%%%%%%%%%%%%%%%%%%%%%%%%%%%%%%%%%%%%%%%%
%%%%%%%%%%%%%%%%%%%%%%%%%%%%%%%%%%%%%%%%%%%%%%%%%

\section{Preliminaries: Information Rate}

Since Shannon’s classical \eco{single-letter} mutual information is \eco{not sufficient} for
channels with memory, information-theoretic analyses of MCvD channels
require evaluating the information rate, which accounts for ISI
memory~\cite{gaye-tez, gaye-achievable}. \eco{The channel information rate between
the transmitted sequence $X^n=(X_1,\ldots,X_n)$ and the molecular count
sequence $Y^n=(Y_1,\ldots,Y_n)$ is defined} as
\begin{equation}
R \triangleq \lim_{n\to\infty}\frac{1}{n} I(X^n;Y^n).
\end{equation}
In MCvD channels, closed-form expressions for $R$ are generally
unavailable due to channel memory and state-dependent noise. \eco{Moreover,
the Rx methods considered in this work do not forward the raw
count sequence $Y^n$, instead, each detector produces a hard
decision sequence
$Z_t \triangleq \hat X_t \in \{0,1\}.$
We therefore evaluate the detector-output information rate
\begin{equation}
R_{\rm det} \triangleq
\lim_{n\to\infty}\frac{1}{n} I(X^n;Z^n),
\label{eq:det_info_rate}
\end{equation}
which quantifies the information after detection. Since the detector output is obtained from
the received observations, the Markov relation
$X^n \!\rightarrow\! Y^n \!\rightarrow\! Z^n$ holds and
$R_{\rm det} \le R$
\label{eq:dpi_detector}
by the data
processing inequality.
The detector-output information rate can be written in terms of entropy
rates. For a stationary discrete process $U=\{U_t\}$, the entropy rate is
defined as
\begin{equation}
\bar H(U) \triangleq
\lim_{n\to\infty}\frac{1}{n}H(U^n).
\end{equation}
Then,
\begin{equation}
R_{\rm det}
=
\bar H(X)+\bar H(Z)-\bar H(X,Z).
\label{eq:entropy_rate_relation}
\end{equation}
For i.i.d. inputs with $\Pr\{X_t=1\}=p_1$, the entropy
rate is
\begin{equation}
\bar H(X)=H_b(p_1)
=
-p_1\log_2 p_1-(1-p_1)\log_2(1-p_1).
\end{equation}
Other entropy rates in \eqref{eq:entropy_rate_relation} can be estimated with context-tree weighting (CTW) estimator \cite{ctw_gao, estimation_information}. Given a simulated sequence
$u^n$ over a finite alphabet and a maximum context depth $D$, CTW assigns
a universal probability $P_{\rm CTW}^{(D)}(u^n)$. The corresponding
entropy-rate estimate is
\begin{equation}
\widehat{\bar H}_{\rm CTW}^{(D)}(U)
=
-\frac{1}{n}\log_2 P_{\rm CTW}^{(D)}(u^n).
\label{eq:ctw_entropy}
\end{equation}
For the detector-output rate, CTW is applied to the binary detector
sequence $z^n$ and to the joint process $(X_t,Z_t)$. The joint process is
encoded over a four-symbol alphabet as $W_t$.
The estimated detector information rate is therefore
\begin{equation}
\widehat R_{\rm det}
=
H_b(p_1)
+
\widehat{\bar H}_{\rm CTW}^{(D)}(Z)
-
\widehat{\bar H}_{\rm CTW}^{(D)}(W).
\label{eq:ctw_detector_rate}
\end{equation}
}
\section{Methodology}

This section details the state-aware methods proposed in this work. Both methods operate on the channel 
with the set of precomputed per-state means/variances $\{\mu_{x,s},\sigma^2_{x,s}\}$.

\subsection{\eco{Causal State Belief}}

\eco{
Let $S_t=(X_{t-1},\ldots,X_{t-m}) \in \{0,1\}^m$ denote the ISI state
at time $t$, and let $f_x(y|s)$ denote the conditional density of $Y_t$
given $X_t=x$ and $S_t=s$. Under the Gaussian approximation in
Section II,
$f_x(y|s)=\mathcal{N}(y;\mu_{x,s},\sigma_{x,s}^2).$
Since the true ISI state is not directly available at the Rx, the
detector maintains a causal belief over the possible ISI states. Before
observing $Y_t$, this belief is defined as
\begin{equation}
\alpha_t(s) \triangleq \Pr\{S_t=s \mid Y^{t-1}\}.
\label{eq:alpha_def}
\end{equation}
Given $\alpha_t$, the predictive density of the next
observation is
\begin{equation}
p(y_t|Y^{t-1})
=
\sum_s \alpha_t(s)
\left[
p_0 f_0(y_t|s)+p_1 f_1(y_t|s)
\right],
\label{eq:predictive_density}
\end{equation}
where $p_x=\Pr\{X_t=x\}$. After observing $Y_t=y_t$, the belief is
updated causally as
\begin{equation}
\alpha_{t\!+\!1}(s')
\propto \!
\sum_s \sum_{x\in\!\{0,1\}}
\!\!\!\alpha_t(s)p_x f_x(y_t|s)
\mathbbm{1}\{s'\!=\!\mathrm{nxt}(s,x)\},
\label{eq:belief_update}
\end{equation}
followed by normalization over $s'$. Here, $\mathrm{nxt}(s,x)$ is the
deterministic next-state function obtained by shifting the current input
bit $x$ into the ISI memory. The proposed BA-MAP and Soft BA-MAP
detectors use this belief to adapt their symbol-wise decision rules to
the current uncertainty over ISI states.
}

\subsection{Soft BA-MAP: Soft-Belief Mixture MAP--LLR Detector}

We perform symbol-wise MAP detection using the full state uncertainty. The Rx maintains a soft belief (forward state distribution) and forms the per-symbol log-likelihood ratio (LLR) via belief-weighted Gaussian mixtures:
%%%%%%%%%%%%%%%%%%%%%%%%%
\begin{align}
\begin{split}
\Lambda_t(y) 
&= \log \frac{p_1 \sum_{s}\alpha_t(s)\,f(y\mid x{=}1,s)}{p_0\sum_{s}\alpha_t(s)\,f(y\mid x{=}0,s)} \\
&= \log \frac{p_1 \sum_{s}\alpha_t(s)\,\mathcal{N}\!\big(y;\mu_{1,s},\sigma^2_{1,s}\big)}
{p_0\sum_{s}\alpha_t(s)\,\mathcal{N}\!\big(y;\mu_{0,s},\sigma^2_{0,s}\big)}.
\end{split}
\label{eq:mixture-llr}
\end{align}
The decision is formulated as $\hat X_t=\mathbbm{1}\{\Lambda_t(Y_t)>0\}$, which minimizes instantaneous bit error under the assumed model that considers ISI and the full mixtures. \eco{Thus, the rule has the form of an LLR, but its weights are causal posterior state beliefs, and its components use state-dependent statistics.}

\subsection{BA-MAP: Belief-Adaptive MAP Thresholding}
To obtain a lightweight decoding mechanism, we replace the full mixtures in~\eqref{eq:mixture-llr} by belief-matched approximations. From the current belief $\alpha_t$, compute the mixture moments
%%%%%%%%%%%%%%%%%%%%%%%%%
\begin{align}
\mu_x^{\text{mix}} &= \sum_s \alpha_t(s)\,\mu_{x,s}, \\
(\sigma_x^{\text{mix}})^2 &= \sum_s \alpha_t(s)\,\big(\sigma^2_{x,s}+\mu_{x,s}^2\big)-\big(\mu_x^{\text{mix}}\big)^2 
\label{eq:mixture-moments}
\end{align}
where $x\in \{0,1\}$, and \eco{Eq.~\eqref{eq:mixture-moments} follows from the law of total variance applied to the belief-weighted mixture.} This method treats $Y_t$ as drawn from two Gaussians with these moments and selects the per-symbol threshold $\tau_t$ by the MAP condition
%%%%%%%%%%%%%%%%%%%%%%%%
\begin{align}
p_0\,\mathcal{N}\!\big(\tau_t;\mu_0^{\text{mix}},(\sigma_0^{\text{mix}})^2\big)
\!=\! p_1\,\mathcal{N}\!\big(\tau_t;\mu_1^{\text{mix}},(\sigma_1^{\text{mix}})^2\big).
\label{eq:map-thr}
\end{align}
For equal variances, $\tau_t$ in \eqref{eq:map-thr} yields the closed form
%%%%%%%%%%%%%%%%%%%%%%%%%%%%
\[
\tau_t=\tfrac{1}{2}\!\left(\mu_0^{\text{mix}}+\mu_1^{\text{mix}}\right)+\frac{(\sigma^{\text{mix}})^2}{\mu_1^{\text{mix}}-\mu_0^{\text{mix}}}\,\log\frac{p_0}{p_1},
\]
otherwise it reduces to a quadratic form with two real roots and we select the root %between the means (or the one 
closest to their midpoint. Although replacing the belief-weighted Gaussian mixture with a single
moment-matched Gaussian introduces approximation error, \eco{the resulting approximation is effective when the belief is concentrated or when the dominant mixture components have similar means and variances. Conversely, under severe ISI at small $T_s$, compressing the full mixture into a single Gaussian inevitably smooths out "peaks and valleys". In the considered settings, BA-MAP remains very close to Soft BA-MAP, suggesting that the moment-matched approximation evaluates state-dependent statistics effectively with a lower computational complexity.}

%%%%%%%%%%%%%%%%%%%%%%%%%%%%%%%%%%%%%%%%%%%%%%%%%
%%%%%%%%%%%%%%%%%%%%%%%%%%%%%%%%%%%%%%%%%%%%%%%%%
%%%%%%%%%%%%%%%%%%%%%%%%%%%%%%%%%%%%%%%%%%%%%%%%%
%%%%%%%%%%%%%%%%%%%%%%%%%%%%%%%%%%%%%%%%%%%%%%%%%

\section{Performance and Complexity Evaluation}\label{sec:performance}

\eco{We evaluate the proposed state-aware signal detectors against standard
equalization and thresholding approaches commonly used in MCvD. The set of
baselines includes Fixed Threshold detection, PA, MMSE equalization,
Genie-MAP, and Viterbi detection. Fixed Threshold applies a single decision
boundary regardless of ISI history and is optimized offline by threshold
sweeping for each $(T_s,\ntx)$.} PA is a Tx-side scheme that adjusts
the emitted molecule count to mitigate average ISI, followed by fixed-threshold
detection. MMSE is a Rx-side linear equalizer that reduces effective ISI
prior to thresholding, \eco{but it introduces decision delay and does not explicitly
use the state-dependent variance. The PA and MMSE baselines are
implemented following~\cite{burcu-MTSK} and~\cite{receiver_akan}. Genie-MAP assumes knowledge of the true ISI state and applies
the corresponding state-dependent MAP decision rule. Thus, it is a genie-aided,
zero-decision-delay reference for state-aware symbol-wise MAP detection. All numerical experiments are performed with a particle-based diffusion simulator. For each parameter setting, $10^6$ transmitted bits are simulated and the molecule-count sequences are result of particle trajectories, while the Gaussian likelihoods described in Section~\ref{section_system_model} are used only inside the analytical detector metrics.}

We briefly analyze the per-symbol complexity of the proposed detectors. For
an ISI memory order $m$, the state space has size $2^m$. Soft BA-MAP evaluates
two belief-weighted Gaussian mixtures, \eco{one for each $x\in\{0,1\}$, and
performs the belief update, and both steps scale as $\mathcal{O}(2^m)$ per symbol.
BA-MAP also scales as $\mathcal{O}(2^m)$ because it computes belief-matched
moments before solving a scalar MAP threshold equation, but it has} lower
constant factors since it avoids explicit mixture-likelihood evaluations.
\eco{Viterbi has comparable state scaling, but additionally requires survivor
storage and traceback delay, while MMSE introduces delay through its linear
equalization structure. In contrast, BA-MAP and Soft BA-MAP make immediate
causal symbol decisions.}

Both proposed detectors store the forward belief vector and therefore require
$\mathcal{O}(2^m)$ memory. In our simulations, $m$ ranges from 5 to 15, and for
typical MCvD parameters the effective memory is often moderate (e.g.,
$m\leq 10$), keeping the proposed belief update and detection steps
computationally feasible. \eco{If stricter implementation constraints are imposed,
a smaller truncated memory order can be used.}

\begin{figure}[!t]
    \centering
    \includegraphics[width=0.97\linewidth]{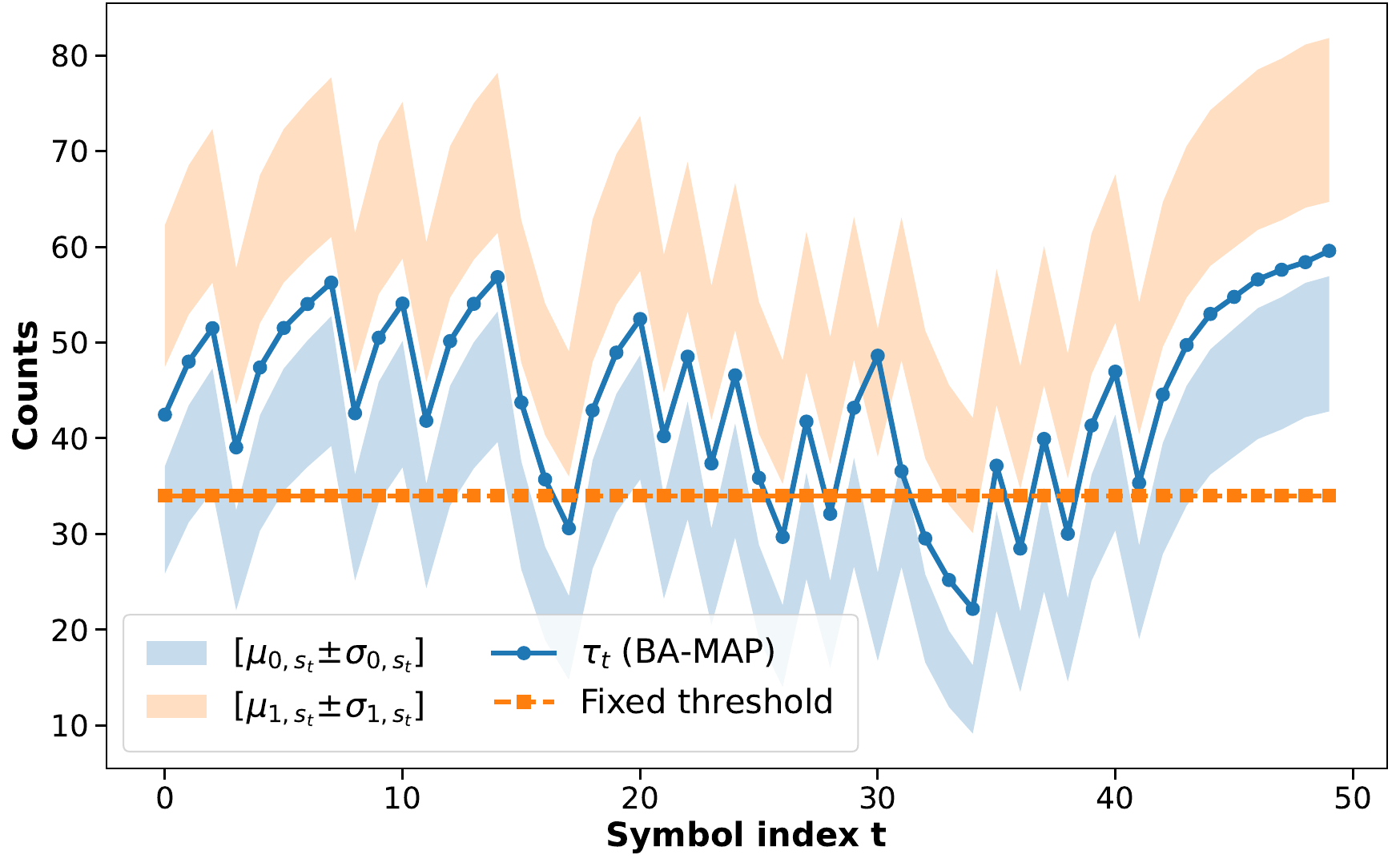}
    \caption{Evolution of true-state Gaussian bands and adaptive thresholds for $N_{\mathrm{Tx}}=250$ over 50 symbols. 
    Shaded regions show the true-state means with $\pm1\sigma$ intervals for bit-0 and bit-1. 
    Solid circular markers indicate the BA-MAP thresholds and square markers show the optimal fixed threshold. 
    The point Tx is located \SI{12.5}{\micro\metre} from the center of a spherical Rx of radius \SI{5}{\micro\metre}, with diffusion coefficient $D=\SI{79.4}{\frac{\micro\metre\squared}{\second}}$ and $T_s = \SI{0.25}{\second}$}.
    \label{fig:bam_thresh}
\end{figure}

Figure~\ref{fig:bam_thresh} illustrates BA-MAP, our state-dependent thresholding mechanism, in a short symbol duration regime. The shaded regions denote the true-state molecule count distributions within one standard deviation around the mean for bit-0 and bit-1 under the current ISI state. In this regime, the received signal power fluctuates rapidly, making a fixed threshold inadequate.%: the horizontal decision line often falls outside the optimal separation region between the two symbol distributions. 
In contrast, the BA-MAP thresholds adapt to the causal state belief by aligning with the belief-matched statistics at every symbol. This adaptivity significantly reduces the error probability compared to a fixed threshold detector.
%%%%%%%%%%%%%%%%%%%%%%%%%%%%%%%%%%%%%%%%%%%%
\begin{figure}[!t]
    \centering
    \includegraphics[width=0.97\linewidth]{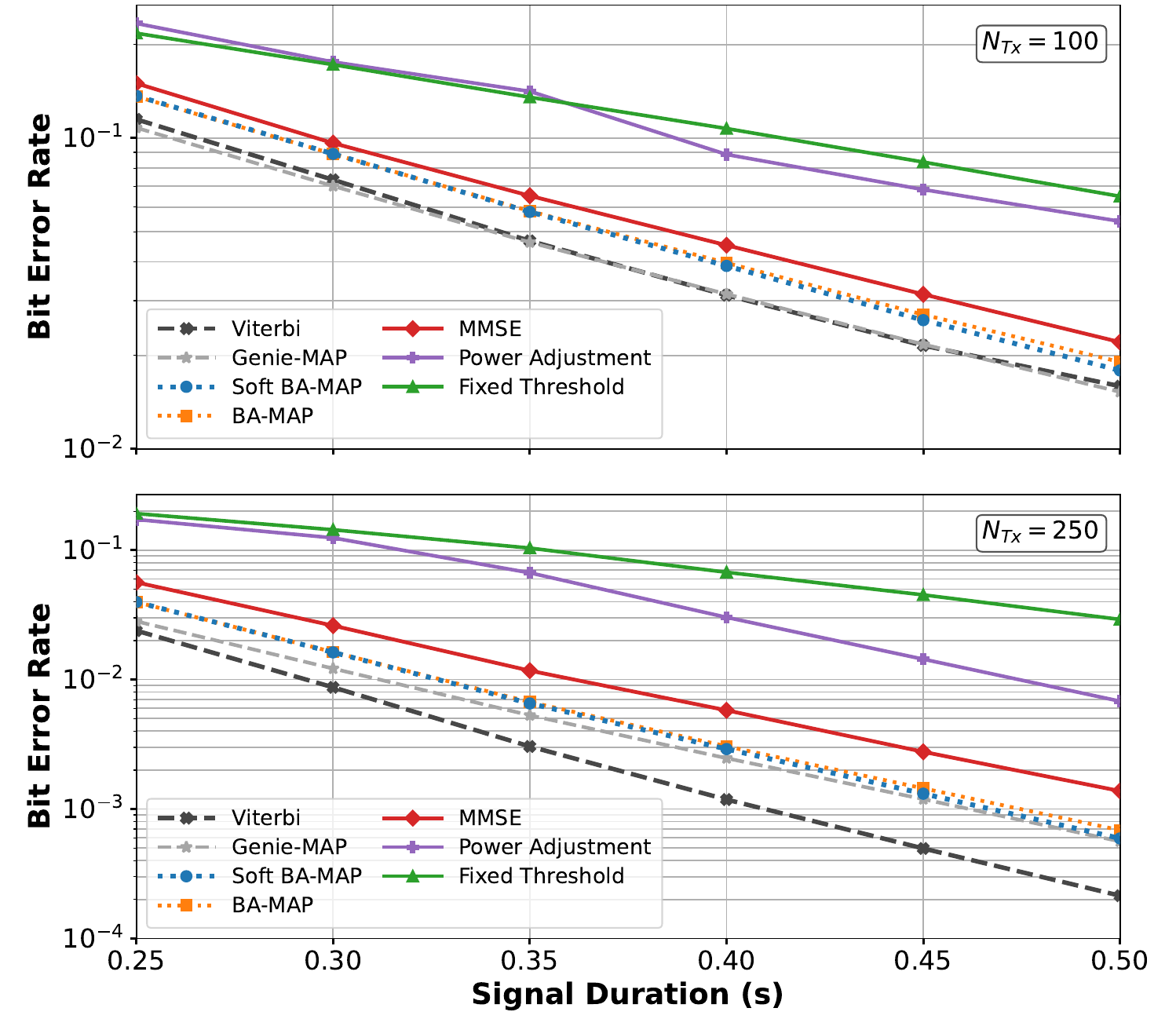}
    \caption{BER performance of the proposed detectors and baseline schemes as a function of symbol duration $T_s$ for average molecule counts $N_{\mathrm{Tx}}=100$ and $N_{\mathrm{Tx}}=250$, using the same channel parameters as in Fig.~\ref{fig:bam_thresh}.}
    \label{fig:ber_side}
\end{figure}

Figure~\ref{fig:ber_side} compares the bit error rates (BER) of different decoding strategies for varying symbol durations and $\ntx$ values. At short symbol durations, heavy ISI dominates, and all methods exhibit higher error rates. Soft BA-MAP achieves the lowest BER \eco{among the practical zero-decision-delay methods}, since it directly incorporates state-dependent mean and variance through mixture likelihoods. \eco{BA-MAP performs close to Soft BA-MAP while maintaining lower complexity by replacing the belief-weighted mixture with a moment-matched Gaussian surrogate. The remaining gap to Genie-MAP is due to the fact that Genie-MAP uses the true ISI state, whereas the proposed detectors rely only on the causal state belief}.
%%%%%%%%%%%%%%%%%%%%%%%%%%%%%%%%%%%%%%%%%%%%%%%%%%%
\begin{figure}[!t] \centering \includegraphics[width=0.95\linewidth]{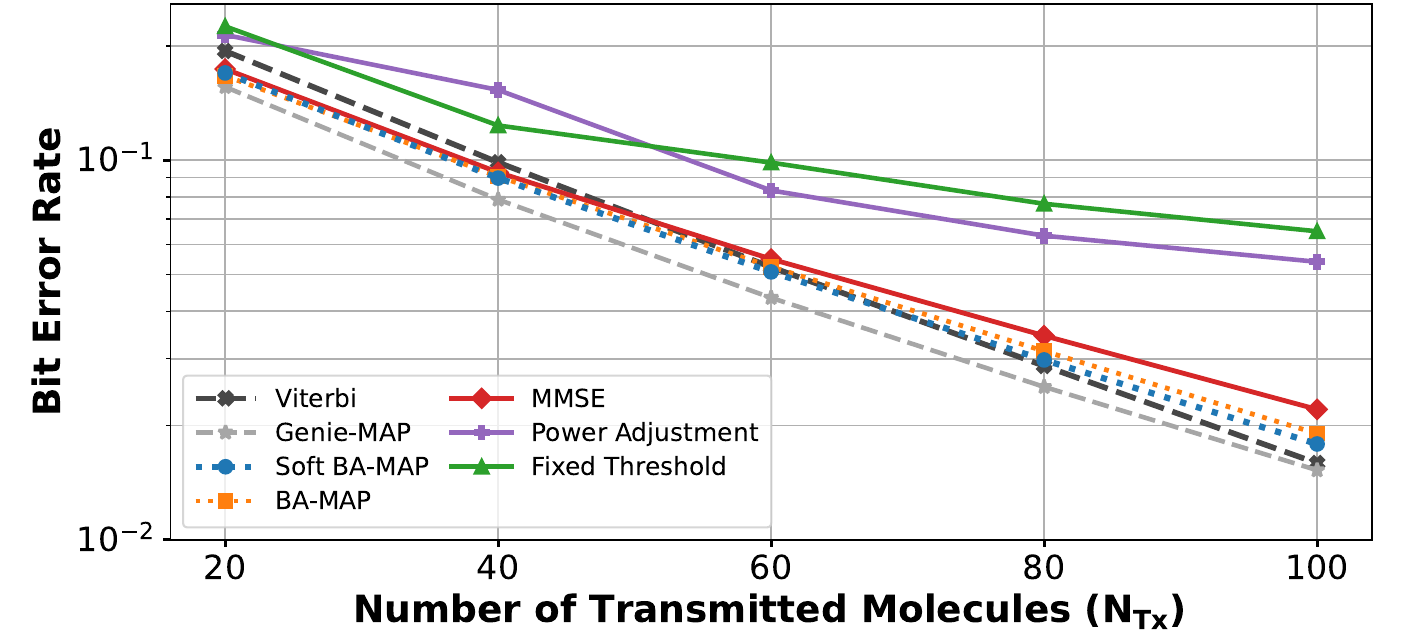} \caption{\eco{BER performance of decoding and equalization strategies versus the number of transmitted molecules $N_{\mathrm{Tx}}$ for $T_s=\SI{0.5}{s}$, using the same channel parameters as in Fig.~\ref{fig:bam_thresh}.}} \label{fig:ntx} 
\end{figure}

\eco{Figure~\ref{fig:ntx} evaluates the detectors with respect to the molecule budget. This sweep includes low-$N_{\mathrm{Tx}}$ regimes where the Gaussian likelihood approximation used by the detectors is expected to be less accurate. Even in this challenging regime, the proposed BA-MAP and Soft BA-MAP detectors preserve their advantage over fixed-threshold, PA, and MMSE baselines, and remain close to the Genie-MAP reference. This indicates that the benefit of exploiting causal ISI-state beliefs and state-dependent count statistics is preserved in the low-count region.}
%%%%%%%%%%%%%%%%%%%%%%%%%%%%%%%%%%%%%%%%%%%%%%%%%%%%%
\begin{figure}[!t]
    \centering
    \includegraphics[width=1.0\linewidth]{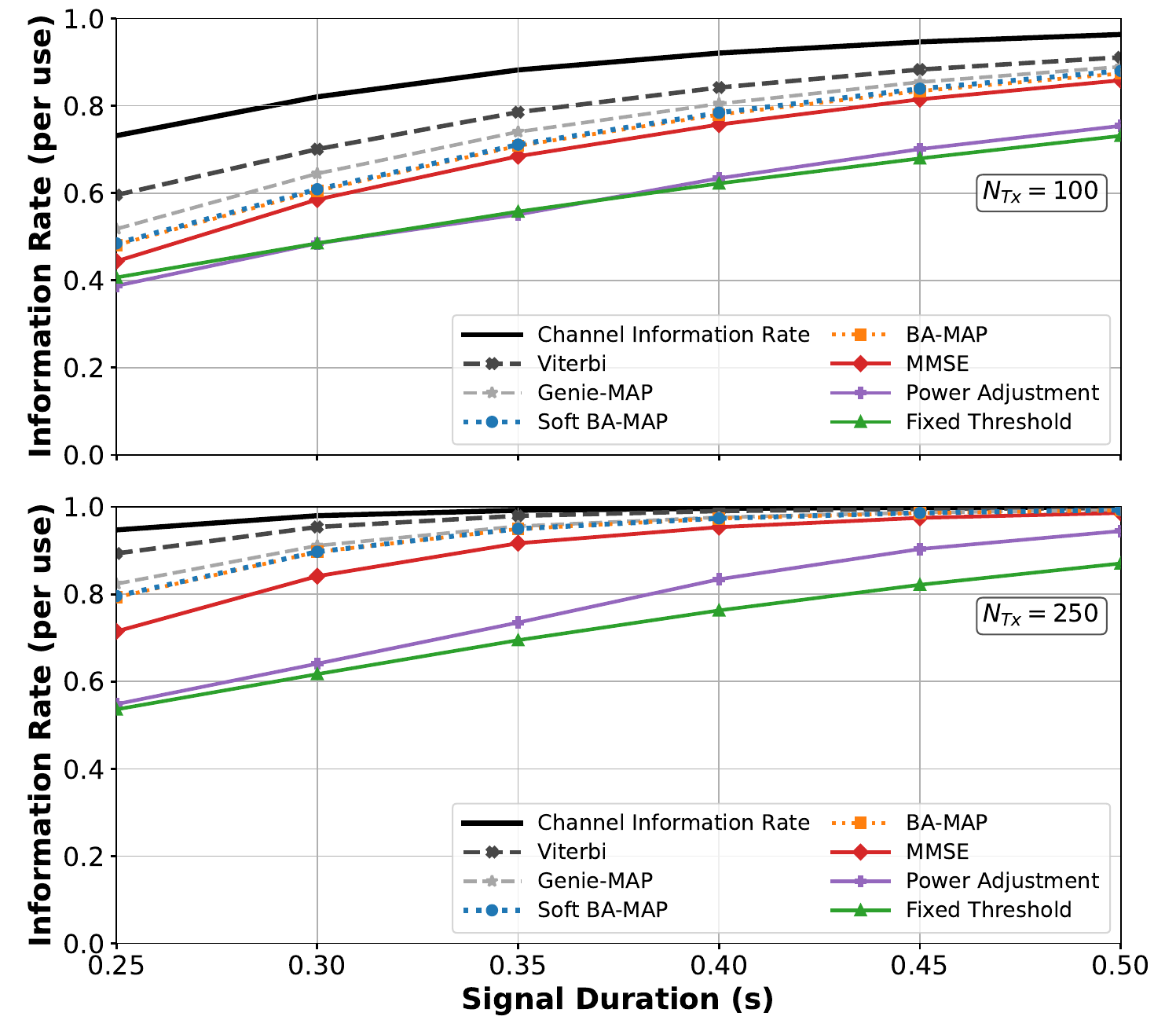}
    \caption{Channel and detector-output information rates of different decoding and equalization strategies with respect to $T_s$, 
    using the same system parameters as in Fig.~\ref{fig:bam_thresh}. \eco{The channel information-rate reference is estimated by the forward information-density estimator, where the predictive density is computed from the causal state belief $\alpha_t(s)$.} }
    \label{fig:rate_side}
\end{figure}

Although the BER curves in Figs.~\ref{fig:ber_side} \eco{and~\ref{fig:ntx}} demonstrate the advantages of the proposed methods, BER alone does not fully capture the true performance of information transfer. \eco{Figure~\ref{fig:rate_side} reports the CTW-based detector-output information rates of all hard-decision methods using the same estimator. The channel information rate is also shown as a reference and upper bounds the detector-output information rates by the data processing inequality. For the information-rate evaluation, the CTW estimates are computed from the same $10^6$-bit Diffusion-simulator output sequences. The only genie-aided baseline is Genie-MAP, which uses the true ISI state when applying the state-dependent MAP decision rule. Hence, the gap between the channel information rate and Genie-MAP reflects the loss caused by mapping the molecular count sequence to a hard-decision sequence. In contrast, the gap between Genie-MAP and the proposed belief-based detectors reflects the residual uncertainty in the causal state belief.}
As shown in Fig.~\ref{fig:rate_side}, \eco{BA-MAP performs very close to Soft BA-MAP despite its lower-complexity Gaussian approximation. Moreover, both proposed detectors operate close to Genie-MAP, showing that the causal belief captures most of the useful state information without requiring true state knowledge. The gain over Fixed Threshold, PA, and MMSE confirms the value of exploiting state-dependent count statistics in hard-output detection. }

\section{Conclusion}

This letter proposed two variance-aware detectors, Soft BA-MAP and BA-MAP, for molecular communication channels impaired by inter-symbol interference (ISI) and heteroscedastic noise. Both methods explicitly model the state-dependent mean and variance of the received molecule counts. Soft BA-MAP uses belief-weighted mixture likelihoods for symbol-wise MAP detection, while BA-MAP replaces these mixtures with moment-matched Gaussian surrogates to obtain adaptive per-symbol MAP thresholds with lower complexity. Numerical results \eco{showed that the proposed detectors reduce BER and improve detector-output information rates compared with conventional fixed-threshold, PA, and MMSE baselines. BA-MAP performs close to Soft BA-MAP with lower complexity, and both methods approach the Genie-MAP reference, indicating that causal state beliefs capture most of the useful ISI-state information without requiring true state knowledge.} Future work will extend the proposed methods to multiple-input multiple-output (MIMO) scenarios. 

\bibliographystyle{IEEEtran} 
\bibliography{refs}

\end{document}